\documentclass[showpacs,amsmath,amssymb,twocolumn,prl,superscriptaddress]{revtex4-1}
\usepackage{amssymb}
\usepackage[dvips]{graphicx}
\usepackage{enumerate}
\usepackage{epsfig}
\usepackage{subfigure}
\usepackage{xcolor}
\usepackage[T1]{fontenc}
\usepackage{fullpage}
\usepackage{amsthm,amsfonts,amssymb,amscd,mathrsfs,xspace,framed}
\usepackage{amsmath}
\usepackage{color}
\usepackage{setspace}
\usepackage{url}
\usepackage{wrapfig}
\usepackage{enumitem}
\newcommand{\MOD}[1]{{{\textcolor{black}{#1}}}}
\begin{document}

\title{\MOD{Adapted} poling to break the nonlinear efficiency limit in nanophotonic lithium niobate waveguides}






\author{Pao-Kang Chen}
\affiliation{J. C. Wyant College of Optical Sciences, University of Arizona, 1630 E. University Boulevard, Tucson, Arizona 85721, USA}
\affiliation{Department of Physics, University of Arizona, 1118 E. Fourth Street, Tucson, Arizona 85719, USA}
\author{Ian Briggs}
\affiliation{J. C. Wyant College of Optical Sciences, University of Arizona, 1630 E. University Boulevard, Tucson, Arizona 85721, USA}
\author{Chaohan Cui}
\affiliation{J. C. Wyant College of Optical Sciences, University of Arizona, 1630 E. University Boulevard, Tucson, Arizona 85721, USA}
\author{Liang Zhang}
\affiliation{J. C. Wyant College of Optical Sciences, University of Arizona, 1630 E. University Boulevard, Tucson, Arizona 85721, USA}
\author{Manav Shah}
\affiliation{J. C. Wyant College of Optical Sciences, University of Arizona, 1630 E. University Boulevard, Tucson, Arizona 85721, USA}
\author{Linran Fan}
\email{lfan@optics.arizona.edu}
\affiliation{J. C. Wyant College of Optical Sciences, University of Arizona, 1630 E. University Boulevard, Tucson, Arizona 85721, USA}

\maketitle

\textbf{
Nonlinear frequency mixing is of critical importance in extending the wavelength range of optical sources~\cite{schliesser2012mid,ghimire2019high,buryak2002optical}. It is also indispensable for emerging applications such as quantum information~\cite{o2009photonic, wehner2018quantum} and photonic signal processing~\cite{he2002optical,cerullo2003ultrafast,langrock2006all}. 
\MOD{Conventional} lithium niobate with periodic poling is the most widely used device for frequency mixing due to the strong second-order nonlinearity. The recent development of nanophotonic lithium niobate waveguides promises improvements of nonlinear efficiencies by orders of magnitude with sub-wavelength optical conferment. 
However, the intrinsic nanoscale inhomogeneity in nanophotonic lithium niobate limits the coherent interaction length, leading to low nonlinear efficiencies~\cite{zhang2022second,chang2016thin,boes2019improved,wang2018ultrahigh,rao2019actively,zhao2020shallow,chen2022ultra}.
Therefore, the performance of nanophotonic lithium niobate waveguides is still far behind \MOD{conventional} counterparts.
Here, we overcome this limitation and demonstrate ultra-efficient second-order nonlinearity in nanophotonic lithium niobate waveguides significantly outperforming \MOD{conventional} crystals~\cite{umeki2010highly,kashiwazaki2020continuous,parameswaran2002highly,parameswaran2002observation,suntsov2021watt,cho2021power,berry2019zn,carpenter2020cw}.
This is realized by developing the \MOD{adapted} poling approach to eliminate the impact of nanoscale inhomogeneity in nanophotonic lithium niobate waveguides. 
We realize overall second-harmonic efficiency near $10^4$\,\%/W without cavity enhancement, which saturates the theoretical limit. Phase-matching bandwidths and temperature tunability are improved through dispersion engineering. The ideal square dependence of the nonlinear efficiency on the waveguide length is recovered.
We also break the trade-off between the energy conversion ratio and pump power. A conversion ratio over 80\,\% is achieved in the single-pass configuration with pump power as low as 20\;mW. 
Our work could provide highly scalable and extremely efficient solutions for next-generation nonlinear optical sources, amplifiers, and converters pivotal for a wide range of classical and quantum applications such as ultra-fast optics and photonic quantum networks.
}

Lithium niobate is one of the most important materials for photonics because of the strong second-order nonlinearity across a broad transparency window~\cite{weis1985lithium}. Therefore, it is widely used in diverse applications ranging from fiber communications to quantum information science~\cite{wooten2000review,elshaari2020hybrid}. Most photonic applications of lithium niobate are based on the electro-optic effect and second-order nonlinear-optical processes. 
Recently, significant efforts have been devoted to the development of the nanophotonic lithium niobate platform. The sub-wavelength confinement of optical fields can significantly increase the nonlinear strength and enable unique capabilities for dispersion engineering and large-scale integration~\cite{zhu2021integrated,boes2023lithium}.

The nanophotonic lithium niobate platform has achieved remarkable advances in applications based on the electro-optic effect. Compared with \MOD{conventional} electro-optic modulators, significant improvements have been demonstrated in almost all key merits including modulation efficiency, bandwidth, long-term stability, and communication data rate~\cite{wang2018integrated,xu2020high,li2020lithium}. Besides the performance improvement, nanophotonic lithium niobate modulators have also been used to demonstrate novel photonic functions such as optical comb generation, non-reciprocal propagation, and quantum transduction~\cite{zhang2019broadband,shah2023visible,xu2021bidirectional}.

\begin{figure*}[t!]
\centering
\includegraphics[width=\textwidth]{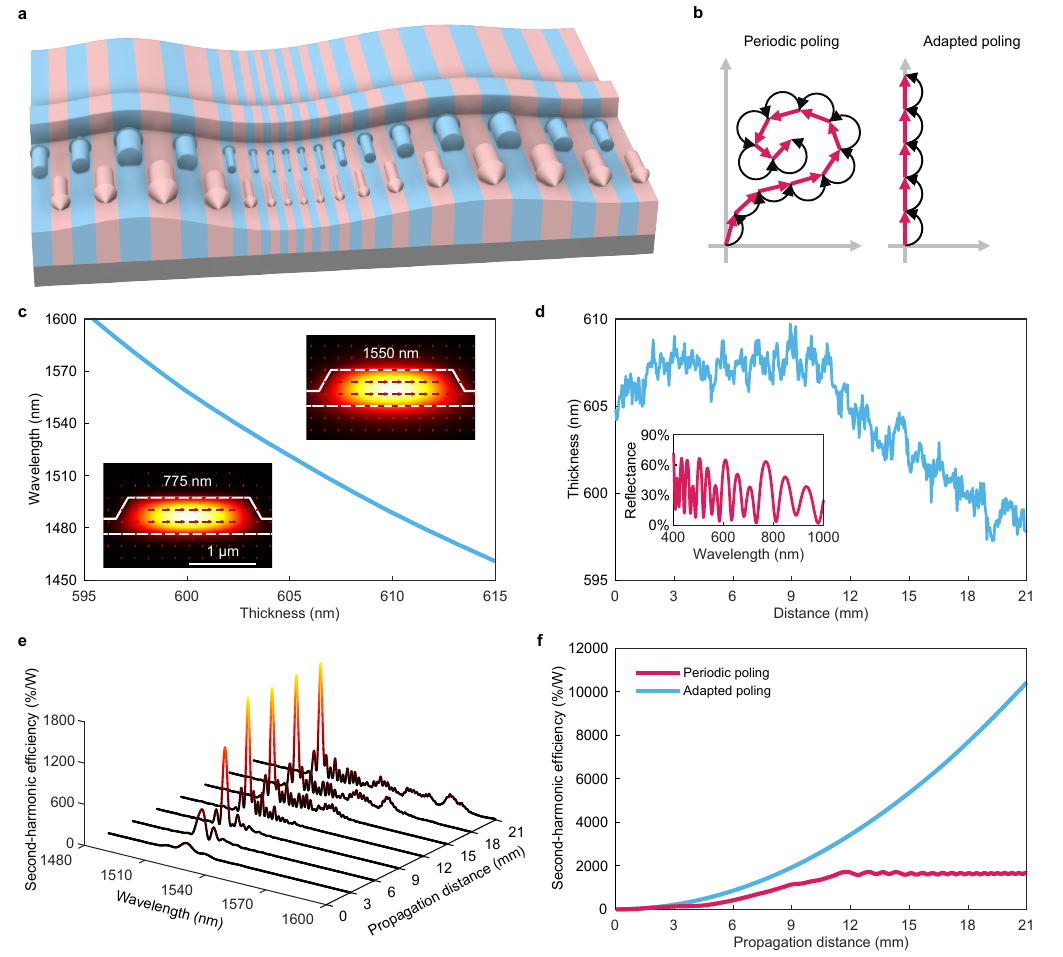}
\caption{\textbf{\MOD{Adapted} poling for nanophotonic lithium niobate waveguides with nanoscale inhomogeneity} \textbf{a,} Device schematic showing the change of poling periods depending on local structure variations. \textbf{b,} Phase diagrams of second-harmonic generation in nanophotonic lithium niobate waveguides with periodic and \MOD{adapted} poling under structure variations, showing random and consistent phase accumulation respectively. \textbf{c,} Simulated phase-matching wavelength for different waveguide thicknesses with a fixed poling period. Inserts: fundamental and second-harmonic mode profiles at 1550\;nm and\;775 nm respectively. \textbf{d,} Measured thickness of the lithium niobate device layer. Inset: measured reflection spectrum for 607.11\;nm lithium niobate layer thickness. \textbf{e,} Calculated second-harmonic spectrum of nanophotonic lithium niobate waveguides with different lengths, assuming periodic poling and thickness variation in \textbf{d}. \textbf{f,} Calculated peak second-harmonic efficiency of nanophotonic lithium niobate waveguides with different lengths using periodic (red) and \MOD{adapted} (blue) poling approaches.
}
\label{Fig1}
\end{figure*}

Similarly, it is expected that the nanophotonic lithium niobate platform can also bring revolutionized performance for applications based on second-order nonlinear-optical processes. Periodically poled lithium niobate waveguides have been the standard device structure for second-order nonlinear-optical processes~\cite{umeki2010highly,kashiwazaki2020continuous,parameswaran2002highly,parameswaran2002observation,suntsov2021watt,cho2021power,berry2019zn,carpenter2020cw}. 
It has been widely demonstrated that nanophotonic lithium niobate waveguides with periodic poling can exhibit nonlinear strength \MOD{(unit\,\%/W/cm$^2$)} orders of magnitude higher than \MOD{conventional} crystals
~\cite{wang2018ultrahigh,rao2019actively,zhao2020shallow,chen2022ultra,chang2016thin,boes2019improved,zhang2022second}. Unfortunately, the increase of the nonlinear strength cannot be translated into the improvement of the nonlinear efficiency \MOD{(unit\,\%/W)}\MOD{, which directly determines the performance of nonlinear processes,} even after intense efforts~\cite{wang2018ultrahigh,rao2019actively,zhao2020shallow,chen2022ultra,chang2016thin,boes2019improved,zhang2022second}.

The intrinsic nanoscale inhomogeneity in nanophotonic lithium niobate limits the coherent interaction length of second-order nonlinear-optical processes to a few millimeters~\cite{wang2018ultrahigh,rao2019actively,zhao2020shallow,chen2022ultra,chang2016thin,boes2019improved,zhang2022second}. \MOD{Currently, the maximum demonstrated nonlinear efficiency in nanophotonic lithium niobate waveguides is around 940\,\%/W~\cite{zhao2020shallow}}. 
\MOD{With much larger coherent interaction length, conventional lithium niobate crystals can achieve nonlinear efficiencies 2400\,\%/W in spite of smaller nonlinear strength~\cite{umeki2010highly}.}
It remains an outstanding question whether nanophotonic lithium niobate waveguides can outperform \MOD{conventional} crystals to achieve efficient second-order nonlinear-optical processes.

\MOD{It is noteworthy that high-efficiency second-order nonlinear processes can be realized in lithium niobate nanophotonic cavities, since the thickness variation is small within the length scale of integrated photonic cavities~\cite{lu2019periodically}. However, waveguide-based second-order nonlinear processes can be highly preferred when large bandwidth, small insertion loss, wide tunability, and high power capability are required.}

Here we overcome this challenge and demonstrate nanophotonic lithium niobate waveguides that outperform \MOD{conventional} crystals in all critical merits including nonlinear efficiency, power consumption, optical bandwidth, and wavelength tunability.
The sub-wavelength confinement of optical fields leads to the increase of the second-order nonlinear strength but also causes the significant sensitivity of the phase-matching condition to structure variations. For example, a thickness change of 10\;nm in the 600-nm lithium niobate device layer can cause a phase-matching wavelength shift over 70\;nm for second-harmonic generation (Fig.\;\ref{Fig1}c). 
Unfortunately, there is significant thickness variation in the nanophotonic lithium niobate device layer caused by intrinsic uncertainties during wafer manufacturing including random ion implantation depth and non-uniform chemical-mechanical polishing rate (Fig.\;\ref{Fig1}d)~\cite{iet:/content/journals/10.1049/el_19950805}.
Non-ideal nanofabrication conditions and poor poling quality further intensify the inhomogeneity.
If the standard period poling technique is used, second-order nonlinear-optical processes will accumulate random phases long the nanophotonic waveguide depending on the local photonic geometry (Fig.\;\ref{Fig1}b). Such random phases prevent the constructive enhancement of second-order nonlinear-optical processes, thus limiting the nonlinear efficiency.
The calculated second-harmonic spectrum is shown in Fig.\;\ref{Fig1}e (See Methods). Significant deviation from the theoretical sinc-squared function can be observed. Instead of the quadratic increase, the peak second-harmonic efficiency starts to saturate with device lengths over 3\;mm and even decreases at long device lengths (Fig.\;\ref{Fig1}f). Therefore, increasing the device length cannot improve the nonlinear efficiency. Currently, efficient second-order nonlinear-optical processes have only been demonstrated in nanophotonic lithium niobate waveguides with device lengths below 5\;mm~\cite{chang2016thin,wang2018ultrahigh,rao2019actively,boes2019improved,zhao2020shallow,chen2022ultra,zhang2022second}.

To compensate for the random perturbation to the phase-matching condition caused by the nanoscale inhomogeneity, we propose the \MOD{adapted} poling approach where the poling period is adjusted depending on the local photonic structure (Fig.\;\ref{Fig1}a). Then, ideal quasi-phase-matching conditions can be realized along the entire waveguide despite the nanoscale inhomogeneity (Fig.\;\ref{Fig1}b). No extra optical loss will be introduced as there is no sudden refractive index change. The constructive enhancement at different waveguide sections leads to the quadratic increase of the second-harmonic efficiency with respect to the device length (Fig.\;\ref{Fig1}f).

\begin{figure*}[t!]
\centering
\includegraphics[width=\textwidth]{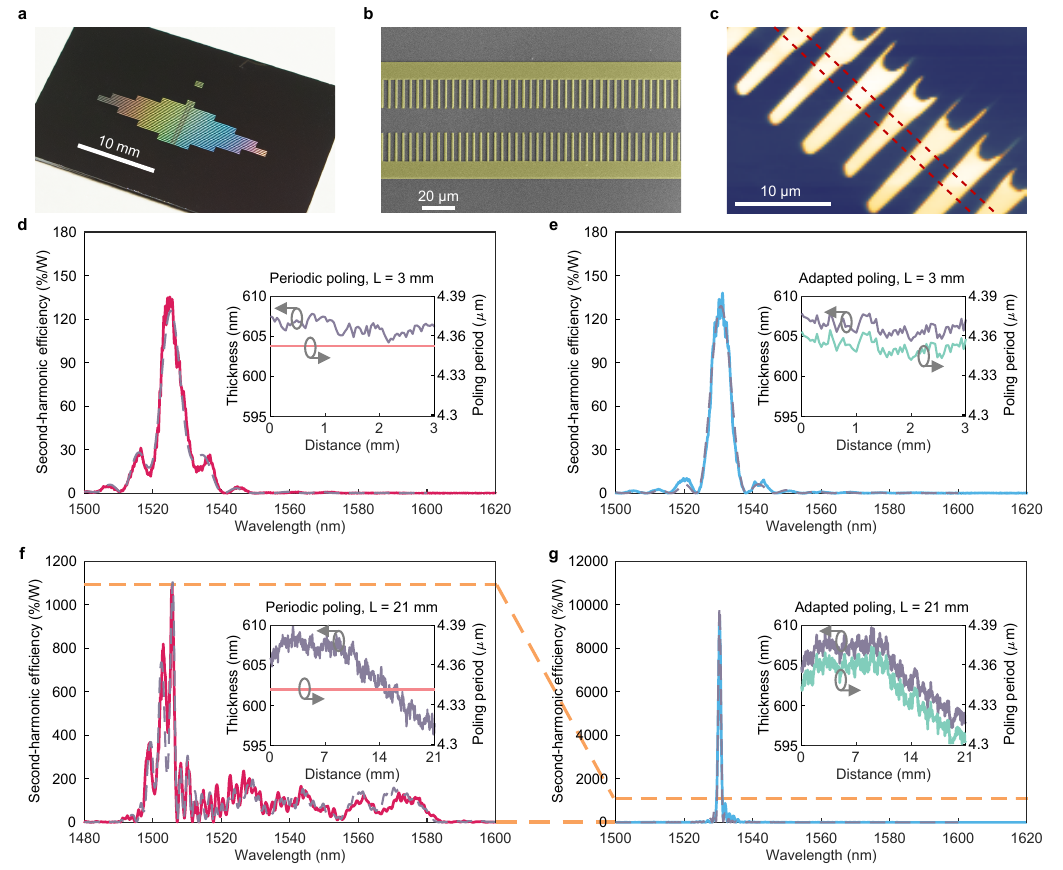}
\caption{\textbf{Performance comparison between periodic and \MOD{adapted} poling approaches.}
\textbf{a,} Optical image of fabricated poling electrodes with different lengths. \textbf{b,} False-color scanning electron microscopy (SEM) image of one poling electrode pair.
\textbf{c,} Piezoresponse force microscopy showing the high-quality domain inversion of lithium niobate. The waveguide position is labeled with dashed lines.
\textbf{d, e, f, g,} measured the second-harmonic generation spectrum of nanophotonic lithium niobate waveguides with periodic poling and 3-mm device length (\textbf{d}), \MOD{adapted} poling and 3-mm device length (\textbf{e}), periodic poling and 21-mm device length (\textbf{f}), and \MOD{adapted} poling and 21-mm device length (\textbf{g}). The measured thickness of the lithium niobate device layer and poling period are shown in the insets. The dashed lines are calculated second-harmonic spectrum using the measured device layer thickness.
}
\label{Fig2}
\end{figure*}

We use X-cut nanophotonic lithium niobate wafers with a 600-nm device layer to fabricate nanophotonic lithium niobate waveguides (See Methods \MOD{and Extended Data Fig.\;1}). The waveguide direction is aligned with the Y-axis of the lithium niobate crystal. We use the fundamental transverse-electric mode for both fundamental and second-harmonic fields around 1550\;nm and 775\;nm respectively (Insets of Fig.\;\ref{Fig1}c). This allows the use of the largest nonlinear coefficient \MOD{$d_{33}=19.5$\;pm/V}. \MOD{It is noteworthy that the nonlinear coefficient changes with the wavelength~\cite{shoji1997absolute}. As the accurate measurement of the nonlinear coefficient for lithium niobate at 1550\;nm has not been performed, we use the value measured at 1313\;nm, which has the smallest wavelength difference with our work~\cite{shoji1997absolute}.}
Before fabrication, we measure the device layer thickness with broadband optical reflectance along the designed waveguide location (Insets of Fig.\;\ref{Fig1}d). Then we pattern the electrode for ferroelectric domain inversion with different poling regime lengths (Fig.\;\ref{Fig2}a\&b). We fabricate two types of nanophotonic lithium niobate waveguides. The first type uses the standard periodic poling approach where the distance between adjacent electrodes fingers is fixed. The second type uses the \MOD{adapted} poling approach where the distance between adjacent electrodes fingers is adjusted based on the measured device layer thickness. High-quality ferroelectric domain inversion with average duty cycle $49\%\pm4\%$ can be achieved by applying optimized high-voltage pulses (Fig.\;\ref{Fig2}c).
Then each waveguide is patterned with multi-pass electron-beam lithography to avoid the waveguide structure distortion at writing field edges.
We also keep sufficient distance between waveguides and chip edges to minimize the influence of the argon-plasma etching non-uniformity. 


With \MOD{3-mm-long waveguides}, both periodic and \MOD{adapted} poling approaches show a second-harmonic spectrum close to the theoretical sinc-squared function (Fig.\;\ref{Fig2}d\&e). Peak efficiencies are also similar with values of 135\,\%/W and 138\,\%/W respectively. This agrees with previous observations that improvements in the second-order nonlinear-optical performance can be achieved with short waveguides~\cite{chang2016thin, wang2018ultrahigh, rao2019actively, boes2019improved, zhao2020shallow, chen2022ultra,  zhang2022second},
as the perturbation to the phase-matching condition caused by the nanoscale inhomogeneity is still relatively small compared with the phase-matching bandwidth.
However, the situation changes dramatically with long waveguides.
We compare the second-harmonic performance of \MOD{21-mm-long} waveguides with periodic and \MOD{adapted} poling approaches. The second-harmonic spectrum with periodic poling deviates significantly from the theoretical sinc-squared function (Fig.\;\ref{Fig2}f). The second-harmonic spectrum shows multiple peaks as different waveguide sections have different phase-matching conditions. The peak efficiency is measured as 1100\,\%/W, which is significantly lower than the theoretical limit of 10500\,\%/W. Using the measured thickness data, we numerically calculate the second-harmonic spectrum only considering the local phase-matching condition change due to the nanoscale inhomogeneity. The strong agreement between calculated and experimental results shows that the nanoscale inhomogeneity is the major factor limiting the overall second-order nonlinear-optical efficiency (Fig.\;\ref{Fig2}f).

With the \MOD{adapted} poling approach, the second-harmonic spectrum still shows the theoretical sinc-squared function. Compared with the \MOD{3-mm-long} waveguides, the second-harmonic spectrum shows much higher peak efficiency and narrower bandwidth (Fig.\;\ref{Fig2}g). This shows that the nanoscale inhomogeneity is compensated by the poling period adjustment along the waveguide. We measure peak efficiency as high as 9500$\pm$1000\,\%/W close to the theoretical limit 10500\,\%/W. 

The bandwidth of the second-harmonic generation is measured as 0.97\;nm in \MOD{the 21-mm-long} waveguides, which is twice larger than \MOD{conventional} lithium niobate crystals with the same length. This is achieved with the smaller group-velocity mismatch between fundamental and second-harmonic fields (\MOD{164}\;fs/mm), enabled by the dispersion engineering capability in nanophotonic lithium niobate waveguides. We further observe that the peak wavelength can be tuned more efficiently in nanophotonic lithium niobate waveguides, reaching 0.2\;nm/$^{\rm o}$C temperature tunability
\MOD{which is better than the temperature tunability of conventional lithium niobate crystals (HC Photonics Corp.) 0.15\;nm/$^{\rm o}$C (See Methods).}

\begin{figure}[t!]
\centering
\includegraphics[width=\linewidth]{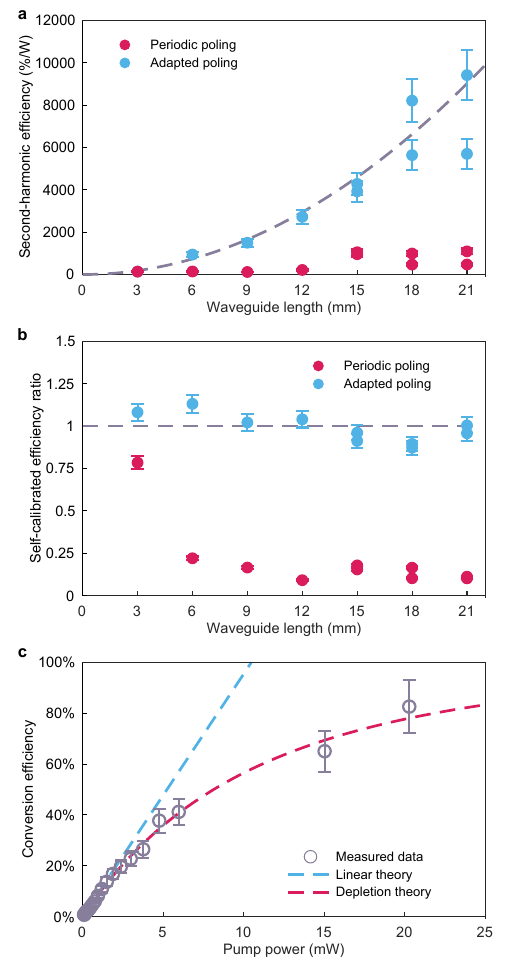}
\caption{
\textbf{Ultra-efficient nanophotonic lithium niobate waveguides.}
\textbf{a,} Peak second-harmonic efficiency with different waveguide lengths. The dashed line shows the square dependence on the waveguide length with a normalized efficiency 2044\,\%/W/cm$^2$ averaged across all waveguides (see Methods).
\textbf{b,} Self-calibrated efficiency ratio characterizing the impact of the nanoscale inhomogeneity. The ideal value (unity) is labeled with the dashed line. 
\textbf{c,} Absolute power conversion efficiency of the \MOD{21-mm-long} waveguide with \MOD{adapted} poling.
}
\label{Fig3}
\end{figure}

Next, we measure the peak efficiency dependence on the waveguide length with both standard periodic poling and \MOD{adapted} poling approaches (Fig.\;\ref{Fig3}a). For the periodic poling approach, the peak efficiency only shows a marginal increase at small waveguide lengths and saturates at large waveguide lengths. In contrast, the peak efficiency with the \MOD{adapted} poling approach follows the quadratic dependence on the waveguide length, matching the theoretical prediction for ideal second-order nonlinear processes in lithium niobate waveguides.

Difficulties in the direct measurement of fiber-waveguide coupling efficiencies and the absolute calibration of power detectors cause the major uncertainty source for the estimation of the second-order nonlinear performance. Here we derive the self-calibrated parameter to characterize the impact of the nanoscale inhomogeneity without requiring absolute power and loss calibration (See Methods). We first \MOD{integrate} the measured second-harmonic efficiency over wavelength to obtain the spectrum area $A$. Then we assume an ideal sinc-squared second-harmonic efficiency spectrum with peak efficiency $\eta{}_0$, and calculate the corresponding spectrum area $A_0$ which is proportional to the ideal peak efficiency $\eta{}_0$. As the nanoscale inhomogeneity conserves the spectrum area up to the dispersion correction, we can use the measured spectrum area $A$ to calibrate the ideal spectrum area $A_0$, thus obtaining the ideal peak efficiency $\eta{}_0$ under the same loss condition with the experiment~\cite{helmfrid1993influence}. Then the ratio between the measured raw second-harmonic peak efficiency ($\eta{}$) and the ideal peak efficiency ($\eta{}_0$) gives the self-calibrated evaluation of the nanoscale inhomogeneity without uncertainties from optical power and loss calibration (Fig.\;\ref{Fig3}b). The efficiency ratio for the \MOD{adapted} poling approach stays constant around unity for all waveguide lengths, while the efficiency ratio for the periodic poling approach drops rapidly to 0.2 when the waveguide length is beyond 6\;mm.

\begin{figure}[tb]
\centering
\includegraphics[width=\linewidth]{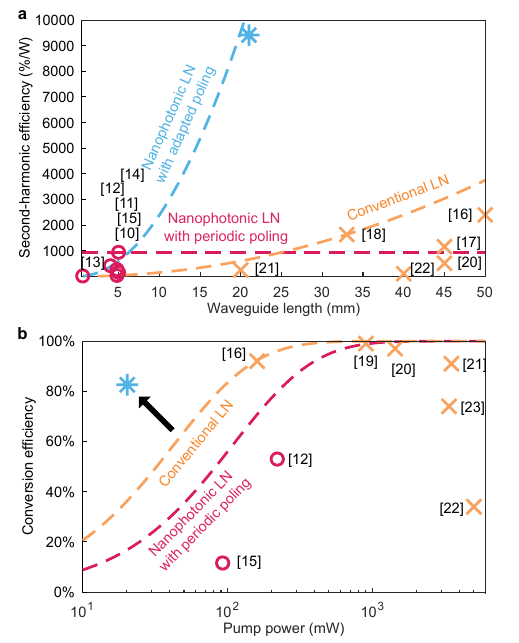}
\caption{
\textbf{Towards ultimate second-order nonlinearity performance.}
Comparison of overall second-harmonic efficiencies with different device lengths (\textbf{a}) and absolution conversion efficiency with different pump powers (\textbf{b}) for this work (blue), previous nanophotonic lithium niobate waveguides (red), and \MOD{conventional} lithium niobate devices (orange). In \textbf{a}, nanophotonic lithium niobate waveguide limit with \MOD{adapted} poling (dashed blue) assumes the theoretical normalized efficiency 2388\,\%/W/cm$^2$, and the \MOD{conventional} limit (dashed orange) assumes the maximum demonstrated value 150\,\%/W/cm$^2$~\cite{parameswaran2002highly}.
In \textbf{b}, \MOD{the state-of-the-art} for nanophotonic lithium niobate waveguides with inhomogeneity (dashed red) and \MOD{conventional} crystals (dashed orange) assume 1000\,\%/W and 2400\,\%/W overall second-harmonic efficiency~\cite{zhao2020shallow,umeki2010highly}. References for the data can be found in Extended Data Table 1.
}
\label{Fig4}
\end{figure}

The optimized second-harmonic generation allows us to realize high absolute conversion efficiency with low pump power. Using the \MOD{21-mm-long} waveguide with \MOD{adapted} poling, we observe the onset of pump depletion and second-harmonic saturation with pump power below 5\;mW. We have measured second-harmonic power of 16.5\;mW with a pump power of 20\;mW, corresponding to 82.5\,\%  absolute power conversion efficiency (Fig.\;\ref{Fig3}c). By fitting the absolute conversion efficiency with the pump-depletion model, we obtain the overall second-harmonic efficiency around 9500$\pm$500\,\%/W, which is consistent with the value measured in the non-depletion region.

We further compare our nanophotonic lithium niobate waveguides using \MOD{adapted} poling with lithium niobate devices using periodic poling in previous works. \MOD{The overall second-harmonic efficiency of our device exceeds both the conventional crystal limit and nanophotonic inhomogeneity limit significantly (Fig.\;\ref{Fig4}a).}
\MOD{Furthermore, we summarize the demonstrated absolute conversion efficiency with both conventional and nanophotonic lithium niobate devices (Fig.\;\ref{Fig4}b).}
In contrast to previous \MOD{conventional} and nanophotonic lithium niobate platforms in which high conversion efficiency can only be achieved with strong pump power, our nanophotonic lithium niobate waveguide with \MOD{adapted} poling breaks the trade-off between conversion efficiency and pump power. We expect that further improvement can be readily made by increasing the length of nanophotonic lithium niobate waveguides. With 4-inch nanophotonic lithium niobate wafers, it is practical to fabricate nanophotonic lithium niobate waveguides with lengths over 90\;mm. Using \MOD{adapted} poling, this can lead to second-harmonic efficiency above $2\times10^5$\,\%/W without cavity enhancement. Pump power below 1\;mW will be sufficient to achieve the same conversion efficiency demonstrated in this work. Even a smaller footprint can be realized by using meandering waveguides.

In summary, we present for the first time an nanophotonic lithium niobate waveguide with overall second-order nonlinear efficiency beyond the nanophotonic inhomogeneous limit and the \MOD{conventional} crystal limit. This is achieved using the \MOD{adapted} poling approach to compensate for the nanoscale inhomogeneity in the nanophotonic lithium niobate platform. The capability to realize ultra-high second-order nonlinear efficiency with an integrated platform is critical for future photonic systems with strong requirements for low power consumption and large-scale manufacturing compatibility. The \MOD{adapted} poling approach can be readily extended to other wavelength ranges and second-order nonlinear interactions. It can also be used to enhance the Kerr nonlinearity through the cascaded second-order nonlinear process~\cite{cui2022situ,cui2022control}. Therefore, this work could have a profound impact on a wide range of nonlinear photonic processes such as optical parametric oscillation and supercontinuum generation in the classical regime, as well as entanglement generation and quantum frequency conversion in the quantum regime. The dramatic performance improvement of these second-order nonlinear processes could open new opportunities for broad applications including ultra-short optical pulse generation, the interconnection between diverse quantum systems, and all-photonic signal processing.


\textbf{Acknowledgments}
This work is supported in part by the U.S. Department of Energy, Office of Advanced Scientific Computing Research (Field Work Proposal ERKJ355); Office of Naval Research (N00014-19-1-2190); NSF-ERC Center for Quantum Networks (EEC-1941583). The reactive ion ether utilized in this study was acquired through an NSF MRI grant (ECCS-1725571). Device fabrication is performed in the OSC cleanroom at the University of Arizona. 

\textbf{Author contributions}
P.C. and L.F. conceived the experiment. P.C. fabricated the device, performed the measurement, analyzed the data, and developed the simulation in useful discussion with C.C. and L.Z.. P.C. and I.B. optimized the poling condition. I.B. and M.S. optimized the fabrication procedure. P.C. and L.F. wrote the manuscript. L.F. supervised the work.

\textbf{Competing interests}
All information in this work is covered by a pending patent application (Application Number 63302331) filed by Pao-Kang Chen and Linran Fan.

\textbf{Data availability}
The data generated and/or analyzed in this work are available from the corresponding author upon reasonable request.

\textbf{Methods}

\textbf{Device fabrication}

\MOD{We first fabricate alignment markers for the thickness measurement, the patterning of electrodes, and the patterning of photonic waveguides.}
After the thickness measurement, nickel electrodes are patterned on top of the lithium niobate device layer. The electrode period is obtained using \MOD{Extended Data Fig.\;2}.
\MOD{The gap between electrodes is 15\;\textmu m. We apply 350\;V electric pulses with 550\;ms duration to nickel electrodes for domain inversion.} Then nickel electrodes are removed by hydrochloric acid. Electron-beam lithography is used to define photonic waveguides with hydrogen silsesquioxane resist. Waveguides patterns are aligned in the poling regime. Then argon-based plasma is used to transfer waveguide patterns into the lithium niobate device layer (600\;nm total thicknesses) with 350\;nm etching depth. The waveguide top width is 1.8\;\textmu m. \MOD{SEM image of the fabricated waveguide is shown in Extended Data Fig.\;1}.



\textbf{Device characterization and measurement uncertainty}

Light is launched into and collected from the waveguide with a pair of lensed fibers. A continuous-wave telecom tunable laser is used as the pump. The polarization is controlled by an in-line fiber polarization controller. Transmitted fundamental and second-harmonic fields are separated with a fiber wavelength-division multiplexer, and detected with InGaAs and Si PIN photodetectors respectively.
Erbium-Doped Fiber Amplifier is used after the telecom laser for high-power measurement, and a band-pass filter is used to remove the amplified spontaneous emission noise.

\MOD{The mean transmission of waveguides is $T_{\rm f,mean}=0.085$ at the fundamental wavelength and $T_{\rm sh,mean}=0.040$ at the second-harmonic wavelength. Fiber-waveguide coupling loss and waveguide propagation loss are both included. The propagation loss is 0.76\;dB/cm at the fundamental wavelength and 0.69\;dB/cm at the second-harmonic wavelength, inferred from on-chip resonator quality factors. With fundamental power $P_{\rm f}$ in the input fiber and second-harmonic power $P_{\rm sh}$ in the output fiber, the nonlinear efficiency for each waveguide is calculated as
\begin{equation}
\begin{aligned}
\eta=\frac{P_{\rm sh}}{\sqrt{T_{\rm sh}}T_{\rm f}P_{\rm f}^2}
\end{aligned}
\end{equation}}
\MOD{The uncertainty is mainly due to the calibration of the transmissions for both fundamental and second-harmonic fields. We estimate the relative uncertainty of the on-chip input fundamental power is $\sim$7\,\%, the uncertainty of the waveguide transmission is $\sim$9\,\%, and the relative uncertainty of the on-chip generated second-harmonic power fluctuation is $\sim$5\,\%. Following the error propagation rule, we calculated the relative uncertainty of the nonlinear efficiency as 12.5\,\%. For the 21-mm-long adapted poling waveguide, this leads to $\pm$1000\,\%/W absolute uncertainty.}

\textbf{Calculation of second-harmonic spectrum}

We use the finite element method to simulate the effective reflective index of both the fundamental field and the second-harmonic field with different waveguide thicknesses, and then calculate the corresponding poling period for the target fundamental wavelength $\lambda$ (Extended Data Fig.\;1). Next, we numerically solve the second-harmonic generation process in the non-depletion regime. We consider $2N$ sections with alternating second-order nonlinearity coefficient
\begin{equation}
\begin{aligned}
\chi_n=\chi_0(-1)^n, n=1,2,...,2N
\end{aligned}
\end{equation}
The section length $L_n$ is half of the poling period. For the standard periodic poling approach, $L_n$ is constant. For the \MOD{adapted} poling approach, $L_n$ depends on the thickness of the $n$th section.

Each section is divided into $M$ cells with equal lengths $L_n/M$. Therefore, the total number of cells is $2N\cdot M$. With input fundamental wavelength $\lambda$, the fundamental and second-harmonic fields in the $m$th cell of the $n$th section ($E_{n,m}^{\rm f} (\lambda)$ and $E_{n,m}^{\rm sh} (\lambda)$) can be calculated from the fields in previous sections.
\begin{equation}
\begin{aligned}
E_{n,m}^{\rm f} (\lambda) =& E_{n,m-1}^{\rm f} (\lambda) \cdot \exp{[ik_{n,m}^{\rm f} (\lambda) L_n/M]} \\
E_{n,m}^{\rm sh} (\lambda) =& E_{n,m-1}^{\rm sh} (\lambda) \cdot \exp{[ik_{n,m}^{\rm sh} (\lambda) L_n/M]}\\
&+\chi_0 (-1)^n [E_{n,m}^{\rm f} (\lambda)]^2 L_n/M
\end{aligned}
\end{equation}

\begin{equation}
\begin{aligned}
E_{n,1}^{\rm f} (\lambda) =& E_{n-1,M}^{\rm f} (\lambda) \cdot \exp{[ik_{n,1}^{\rm f} (\lambda) L_n/M]} \\
E_{n,1}^{\rm sh} (\lambda) =& E_{n-1,M}^{\rm sh} (\lambda) \cdot \exp{[ik_{n,1}^{\rm sh} (\lambda) L_n/M]}\\
&+\chi_0 (-1)^n [E_{n,1}^{\rm f} (\lambda)]^2 L_n/M \\
\end{aligned}
\end{equation}
with $k_{n,m}^{\rm f}$ and $k_{n,m}^{\rm sh}$ the wavevectors for the fundamental and second-harmonic fields respectively. Initial conditions are set as
\begin{equation}
\begin{aligned}
E_{1,1}^{\rm f} (\lambda) =& E_0 \\
E_{1,1}^{\rm sh} (\lambda) =& 0 \\
\end{aligned}
\end{equation}
Then the overall second-harmonic efficiency at wavelength $\lambda$ is given by
\begin{equation}
\begin{aligned}
\eta (\lambda)\propto \big|E_{N,M}^{\rm sh}(\lambda)\big|^2/|E_0|^4
\end{aligned}
\end{equation}
In this calculation, we assume low pump power therefore the second-harmonic generation is in a non-depletion regime. We also ignore the change of the second-order nonlinear coefficient ($\chi_0$) due to the optical mode size. Therefore, only the local phase-matching condition change due to the nanoscale inhomogeneity is included.




\textbf{Self-calibrated second-harmonic generation performance}

The area of the second-harmonic efficiency spectrum is calculated as
\begin{equation}
\begin{aligned}
    A=\int \eta(\lambda) \,d\lambda
\end{aligned}
\end{equation}
where $\eta(\lambda)$ is the measured second-harmonic efficiency. The measured peak efficiency is $\eta=\max[\eta(\lambda)]$. We can also calculate the spectrum area assuming the ideal phase-matching condition along the entire waveguide~\cite{helmfrid1993influence}
\begin{equation}
\begin{aligned}
    A_0&=\alpha\eta_0/L
\end{aligned}
\end{equation}
where $\eta_0$ is the peak efficiency for the ideal second-harmonic spectrum and $\alpha=2\pi(d\Delta k/d\lambda)^{-1}$ is the dispersion factor with $\Delta k$ the wavevector mismatch between fundamental and second-harmonic fields. Under the same loss condition, the spectrum area is conserved under inhomogeneous broadening~\cite{helmfrid1993influence}.
\begin{equation}
\begin{aligned}
    A&=A_0
\end{aligned}
\end{equation}
Therefore, we can get the ideal peak efficiency $\eta_0$ under the same loss condition, but with the ideal phase-matching condition along the entire waveguide.
\begin{equation}
\begin{aligned}
    \eta_0&=AL/\alpha
\end{aligned}
\label{eta_0}
\end{equation}
Then we can have the self-calibrated efficiency ratio between the measured and ideal peak efficiencies
\begin{equation}
\begin{aligned}
    R=\frac{\eta}{\eta_0}=\frac{\eta \alpha}{AL}
\end{aligned}
\end{equation}
Assuming uncalibrated extra losses $t_1$ and $t_2$ for the input fundamental and output second-harmonic fields respectively, we have
\begin{equation}
\begin{aligned}
    \frac{\eta}{A}=\frac{\eta' t_2t_1^{-2}}{\int \eta'(\lambda)t_2t_1^{-2} \,d\lambda}
    =\frac{\eta'}{A'}
\end{aligned}
\label{loss}
\end{equation}
where $\eta'$ and $A'$ are the second-harmonic efficiency and spectrum area with perfect loss calibration. Eq.\;(\ref{loss}) shows that the efficiency ratio $R$ is independent of loss and power calibration. 

Using the ideal peak efficiency $\eta_0$ (Eq.\;\ref{eta_0}), we can also obtain the intrinsic normalized efficiency without the impact of nanoscale inhomogeneity
\begin{equation}
\begin{aligned}
   \eta_{0,norm}=\eta_0/L^{2}=A/L\alpha
\end{aligned}
\end{equation}
The average intrinsic normalized efficiency across all waveguides is 2044$\pm$150\,\%/W/cm$^2$. This value is slightly smaller than the theoretical normalized efficiency 2388\,\%/W/cm$^2$ mainly due to the deviation of poling duty cycles from 50\%.

\textbf{Temperature tunability}

The phase-matching condition can be written as
\begin{equation}
\begin{aligned}
   \frac{\pi}{\Lambda}=\frac{2\pi (n_2-n_1)}{\lambda}
\end{aligned}
\label{PMcond}
\end{equation}
where the poling period $\Lambda$ depends on temperature $T$ due to thermal expansion, and fundamental and second-harmonic effective refractive indices ($n_1$ and $n_2$) depend both on the temperature $T$ and wavelength $\lambda$. 
By differentiating Eq.\;(\ref{PMcond}), we obtain the temperature tunability
\begin{equation}
\begin{aligned}
\frac{d\lambda}{dT}=\frac{\frac{\partial \Lambda}{\partial T}(n_2-n_1)+(\frac{\partial n_2}{\partial T}-\frac{\partial n_1}{\partial T})\Lambda}
{\frac{1}{2}-(\frac{\partial n_2}{\partial \lambda} - \frac{\partial n_1}{\partial \lambda})\Lambda}
\end{aligned}
\end{equation}
The simulated temperature tunability of nanophotonic lithium niobate waveguides is 0.34\;nm/$^{\rm o}$C, in contrast to 0.13 nm/$^{\rm o}$C for \MOD{conventional} lithium niobate. This is mainly attributed to dispersion control. Instead of a constant negative dispersion value in \MOD{conventional} lithium niobate ($(\frac{\partial n_2}{\partial \lambda} - \frac{\partial n_1}{\partial \lambda}) \Lambda=\MOD{-0.608}$), integrate waveguides show positive dispersion value ($(\frac{\partial n_2}{\partial \lambda} - \frac{\partial n_1}{\partial \lambda}) \Lambda=\MOD{0.363}$), boosting the temperature tunability \MOD{(Extended Data Fig.\;3 and 4)}.







\bibliography{maincitation}

\end{document}